\documentclass[twocolumn,floatfix,aps,prb]{revtex4}
\usepackage{graphicx}
\usepackage{dcolumn}
\begin{document}

\title{Ferromagnetic to spin glass cross over in
(La,Tb)$_{2/3}$Ca$_{1/3}$MnO$_3$}

\author{C. Raj Sankar}
\author{S. Vijayanand}
\author{P. A. Joy}
\email{pa.joy@ncl.res.in}
\affiliation{Physical and Materials Chemistry Division, National
Chemical Laboratory, Pune 411008, India}

\begin{abstract}
In the series La$_{2/3-x}$Tb$_x$Ca$_{1/3}$MnO$_3$, it is known that the
compositions are ferromagnetic for smaller values of $x$ and show spin
glass characteristics at larger values of $x$. Our studies on the
magnetic properties of various compositions in the
La$_{2/3-x}$Tb$_x$Ca$_{1/3}$MnO$_3$ series show that the cross over
from ferromagnetic to spin glass region takes place above $x \approx$
1/8. Also, a low temperature anomaly at 30 K, observed in the ac
susceptibility curves, disappears for compositions above this critical
value of $x$. A mixed phase region coexists in the narrow compositional
range 0.1 $\leq x \leq$ 0.125, indicating that the ferromagnetic to
spin glass  cross over is not abrupt.

\end{abstract}

\maketitle

\section{Introduction}

There has been a wide interest in the study of the substituted
perovskite-type manganites, La$_{1-x}$A$_x$MnO$_3$, in order to
understand the different aspects of the complex magnetic behaviour
exhibited by these compounds.\cite{tok00,dag03,boris04} The double
exchange interactions involving Mn$^{3+}$ and Mn$^{4+}$ ions give rise
to ferromagnetism in the substituted manganites. The changes in the
Mn-O-Mn bond angle, from structural distortions, are very crucial in
determining the strength of the ferromagnetic interactions. Highest
Curie temperature in the La$_{1-x}$Ca$_x$MnO$_3$ series is obtained for
$x \approx$ 1/3 and hence there are many studies reported for the
composition La$_{2/3}$A$_{1/3}$MnO$_3$.  Many interesting new magnetic
behaviors are observed when La$^{3+}$ is partially substituted by other
trivalent rare-earth ions in La$_{2/3}$Ca$_{1/3}$MnO$_3$,
\cite{hwa95,mai96,bla96,det96,ter98,nie99} though there are no changes
in the Mn$^{3}$/Mn$^{4+}$ ratio after substitution. Hwang {\itshape et
al.}\cite{hwa95} found a direct correlation between the Curie
temperature and the average ionic radius of the La-site ions, where the
Curie temperature decreases with decreasing average ionic radius in
La$_{0.7-x}$R$_x$Ca$_{0.3}$MnO$_3$ (R = Pr, Y), indicating the role of
lattice effects in determining the ferromagnetic properties. Partial
substitution of La by smaller ions leads to a decrease in the Mn-O-Mn
bond angle and this affects the long range ferromagnetic exchange
interactions.\cite{bla96,det96,ter98}

For La$_{0.67-x}$Tb$_x$Ca$_{0.33}$MnO$_3$ (in the reports, the chemical
composition is used as (La$_{1-x}$Tb$_x$)$_{2/3}$Ca$_{1/3}$MnO$_3$), a
gradual decrease in the Curie temperature is observed with increasing
concentration of Tb.\cite{det96,nie99} This is also associated with a
broadening of the magnetic transition and ultimately a spin glass
behaviour is observed at larger concentrations of Tb.
\cite{det97,wat01}  In La$_{0.67-x}$Tb$_x$Ca$_{0.33}$MnO$_3$, the
evolution of a spin glass (SG) or a cluster glass (CG) state is thought
to be due to the competing interactions of ferromagnetic (FM) and
antiferromagnetic (AFM) types or the random distribution of Mn-O-Mn
bond angle which suppresses the exchange strength between the Mn ions
significantly.\cite{det96,ter98,wat01} At sufficiently large values of
$x$ ($x$ = 0.22), short range ordered magnetic clusters are formed with
typical magnetic coherence length of around 18 \AA, at low
temperatures.\cite{det97}

The reported studies on La$_{0.67-x}$Tb$_x$Ca$_{0.33}$MnO$_3$  are
performed on ferromagnetic compositions ($x \leq$ 0.1) or spin glass
compositions ($x >$ 0.15) and therefore no information is available
on the changeover region from ferromagnetic to spin glass
characteristics. In order to explore the  critical concentration of
the Tb ions which results in the formation of magnetic clusters
leading to spin glass characteristics, a series of compositions with
very close values of $x$ between 0.1 and 0.15 in
La$_{0.67-x}$Tb$_x$Ca$_{0.33}$MnO$_3$ have been studied. The
critical concentration is found to be $x \approx$ 1/8, and
interesting magnetic properties are observed for compositions close
to the cross over region.

\section{Experimental}
The polycrystalline La$_{0.67-x}$Tb$_x$Ca$_{0.33}$MnO$_3$ compositions
were synthesized by the conventional solid state route from
La$_2$O$_3$, Tb$_4$O$_7$, CaCO$_3$ and MnO$_2$ by mixing these oxides
in the required stoichiometry for $x$ = 0, 0.03, 0.07, 0.10, 0.11,
0.12, 0.125, 0.13, 0.15, 0.20, and 0.25. The well-mixed powders were
initially heated at 1273 K for 48 h, and subsequently at 1473 K for
48h, 1573 K for 48 h and finally at 1623 K for 24h, with intermediate
grindings at every 24 h steps to ensure the sample homogeneity. Finally
the powder samples were pelletized and sintered at 1673 K for 24 h. The
samples were characterized by powder X-ray diffraction  using Cu
K$\alpha$ radiation and Ni filter. The Mn$^{4+}$ contents in the
compounds were estimated by the iodometric titration
method.\cite{san05} The magnetization measurements were performed on a
vibrating sample magnetometer. Temperature variation of the ac magnetic
susceptibility of the samples was measured by the mutual inductance
method in a field of 2 Oe and at a frequency of 210 Hz.

\section{Results and Discussion}

\begin{figure}
\includegraphics[width=3in,height=2.5in]{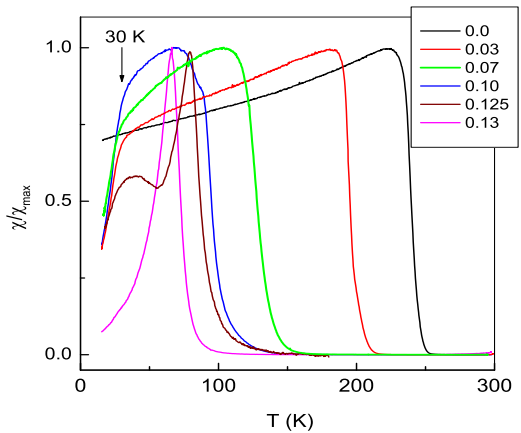}
\caption{\label{fig1} Temperature variation of the ac susceptibility
of some compositions in La$_{0.67-x}$Tb$_x$Ca$_{0.33}$MnO$_3$. The
numbers indicate the values of $x$.}
\end{figure}

The samples were characterized for their phase purity by X-ray
diffraction. All samples showed single phase nature and the diffraction
patterns were indexed on the distorted orthorhombic structure with
space group \emph{Pbnm}.\cite{bla96} The lattice parameters were
obtained by least squares refinement of the diffraction patterns and
found to be comparable for those of the corresponding compositions
reported previously.\cite{bla96,wat01} The Mn$^{4+}$ content was found
to be matching with the expected value of 33\% for all the
compositions.

The ac susceptibility curves of the unsubstituted and some
Tb-substituted compositions are shown in Fig.~\ref{fig1}. With
increasing concentration of Tb, the magnetic transition temperature is
decreased, and the susceptibility curves show a cusp-like feature when
the concentration is increased beyond $x$ = 0.1.  The shapes and
features of the ac susceptibility curves of the Tb-substituted
compositions (for $x \leq$ 0.1 and $x >$ 0.15) are similar to that
reported earlier.\cite{bla96,nie99,det97} Although well-defined
magnetic transitions are observed for the samples with low
concentrations of Tb, there is a distinct anomalous feature in the
susceptibility curves of these compositions at low-temperatures. A
decrease in the susceptibility is observed below $\sim$30 K, as soon as
a small amount of Tb is incorporated ($x$ = 0.03).

\begin{figure}
\includegraphics[width=3in,height=2.5in]{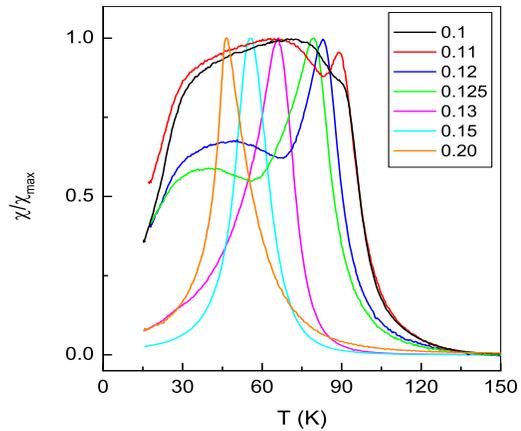}
\caption{\label{fig2} Temperature dependence of the ac
susceptibility of the compositions in
La$_{0.67-x}$Tb$_x$Ca$_{0.33}$MnO$_3$, for 0.1 $\leq x \leq$ 0.2.
The numbers indicate the values of $x$.}
\end{figure}

As shown in Fig.~\ref{fig2}, for $x$ = 0.1,  a small step in the
susceptibility curve is observed below the main magnetic transition.
For $x$ = 0.11, there is a cusp like feature observed at a higher
temperature which is succeeded by the normal flat curve, and finally a
drop in the susceptibility below 30 K. This is a complex feature and is
not commonly reported for the manganite compositions. As the
concentration of Tb ions is further increased, it can be seen that the
cusp-like feature becomes more prominent, the flat region gradually
vanishes, but a dip in susceptibility is still observed at 30 K for the
compositions up to $x$ = 0.125. For $x >$ 0.125 there is only a
cusp-like feature in the ac susceptibility curve as observed for spin
glass systems. The composition with $x$ = 0.33 is reported to show spin
glass behavior.\cite{det96} Neutron diffraction studies have indicated
ferromagnetically ordered state for $x$ = 0.067 at low temperatures,
whereas short range ordered clusters are found for $x$ = 0.2 (the
authors have used the chemical formula as
(La$_{1-x}$Tb$_x$)$_{2/3}$Ca$_{1/3}$MnO$_3$, so that $x$ = 0.1 and 0.3,
respectively, correspond to $x$ = 0.067 and 0.2 in
La$_{0.67-x}$Tb$_x$Ca$_{0.33}$MnO$_3$).\cite{wat01} These clusters show
spin glass-type of behavior due to competition between FM and AFM types
of exchanges.\cite{bla96,nie99}

For the compositions with $x >$ 0.125, there is a  gradual shift
of the cusp towards lower temperatures as the value of $x$ is
increased. Also, it was found that the relative susceptibility
value also decreased with the increasing Tb concentration. These
observations point out  the  evolution of short range ordering due
to the decreasing Mn-O-Mn bond angle and consequent formation of
magnetic clusters as the concentration of Tb is increased. This
clustering is likely to be due to the breaking or weakening of the
double exchange closer to the Tb-sites caused by the local
structural distortion  and the spin glass like feature originates
from the formation of such clusters. The decreasing temperature
corresponding to the cusp, identified as T$_g$ henceforth,
suggests the confinement of magnetic clusters to shorter length
scales as the number of Tb centers is increased (increasing values
of $x$).

\begin{figure}
\includegraphics[width=3in,height=2.5in]{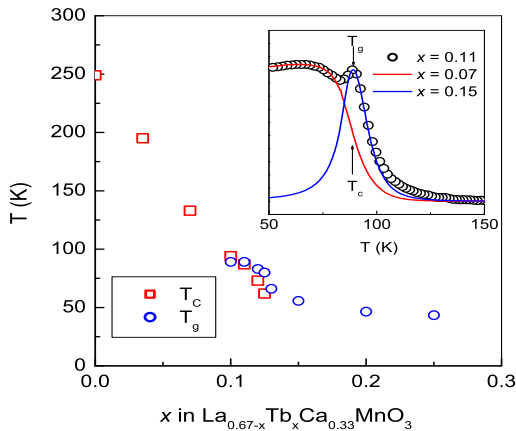}
\caption{\label{fig3} Variation of $T_c$ and $T_g$ as a function of
$x$ in La$_{0.67-x}$Tb$_x$Ca$_{0.33}$MnO$_3$.  Inset: illustration
of the method of extracting $T_c$ and $T_g$, for $x$ = 0.11, using
the data for $x$ = 0.15 (for $T_g$) and $x$ = 0.07 (for $T_c$).}
\end{figure}

As shown in Fig.~\ref{fig3}, the ferromagnetic transition temperature
decreases almost linearly with increasing $x$, up to $x$ = 0.125. For
the compositions showing both ferromagnetic and spin glass
characteristics (0.1 $\leq x \leq$ 0.125), $T_c$ and $T_g$ are
determined as illustrated in the inset of Fig.~\ref{fig3} for $x$ =
0.11. Here, the $\chi$-T data for $x$ = 0.15 (spin glass composition)
is shifted towards the right side and that of $x$ = 0.07 (ferromagnetic
composition) is shifted towards the left side along the $x$-axis, to
match with the observed data of the mixed phase composition and the
curves are normalized with respect to the maximum values. $T_c$ is
taken as the mid point of the magnetic transition.  T$_g$ changes
abruptly around $x$ = 0.125 and varies from 66 K for $x$ = 0.13 to 43 K
for $x$ = 0.25. $T_g$ has been found to be almost independent (40-50 K)
of $x$ for higher Tb concentrations.\cite{bla96} Thus, the lowest value
of possible $T_g$ is larger than the temperature (30 K) where a
decrease in the ac susceptibility is observed for 0 $ < x \leq$ 0.125.

The decrease in the susceptibility below 30 K as soon as a small amount
of Tb is incorporated in the lattice of La$_{2/3}$Ca$_{1/3}$MnO$_3$ has
been ascribed to spin glass behavior.\cite{nie99} However,  neutron
diffraction studies showed ferromagnetic ordering for $x$ = 0.067 down
to 7 K.\cite{wat01} Thus, the feature at 30 K for 0.03 $\leq x \leq$
0.125 is not likely to be spin glass transition. It is possible that at
small concentrations the Tb$^{3+}$ ions are randomly distributed in the
lattice, the double exchange is disturbed around the Tb centers and
therefore tiny magnetic clusters are formed with reduced Mn-O-Mn angle.
These tiny magnetic clusters remain isolated until $x$ = 0.125 (1/8)
above which larger magnetic clusters are formed due to the breaking or
considerable weakening of the three dimensional long range ordering.
Thus, the temperature at which a decrease in the susceptibility is
observed, due to these small clusters, remains the same until $x$ =
1/8. The three dimensional ordering is affected when $x>$ 1/8. However,
there is another possibility that, at small concentrations, the Tb ions
form a TbMnO$_3$ like local environment in the lattice of
La$_{0.67-x}$Tb$_x$Ca$_{0.33}$MnO$_3$. For TbMnO$_3$, an
incommensurate-commensurate phase transition, which is accompanied by a
ferroelectric transition, associated with a lattice modulation, is
observed close to $\sim$30 K and large magnetic field controlled
polarization effects are reported at this temperature.
\cite{kim03,got04}

A local phase separation exists in the $x$ = 0.1 sample, as
evidenced by a small step- like magnetic transition, indicating that
 ferromagnetic clusters are started forming at this value of
$x$. It is possible that, above this value of $x$, some magnetic
clusters with short range ordering are separated whose size
decreases with increasing $x$. Thus, at intermediate  values of $x$,
the lattice is consisting of larger ferromagnetic clusters with
sufficiently long range ordering and smaller short range ordered
clusters. Also, for the larger long range ordered part, the magnetic
transition temperature decreases due to the decrease in the Mn-O-Mn
bond angle. Thus, the clustering may be seen to start when $x =$
0.11, where a cusp is also observed along with the normal magnetic
features.

\begin{figure}
\includegraphics[width=3in,height=2.5in]{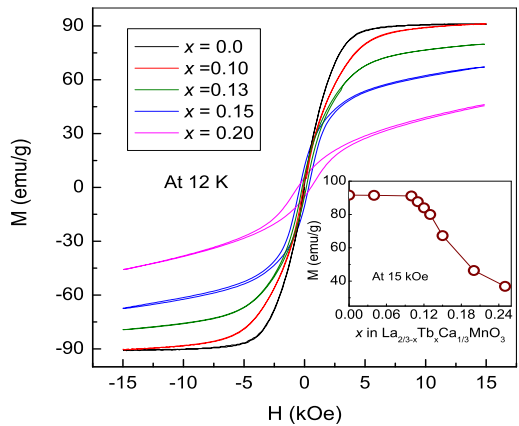}
\caption{\label{fig4} The M-H curves of
La$_{0.67-x}$Tb$_x$Ca$_{0.33}$MnO$_3$ for different values of $x$.
Inset: variation of magnetization at 15 kOe as a function of $x$.}
\end{figure}

Fig.~\ref{fig4} shows the dc magnetization curves of different
compositions, measured at 12 K, up to a maximum field of 15 kOe. The
magnetization is saturated above 10 kOe for $x \leq$ 0.1. The variation
of the magnetization at the maximum measured field of 15 kOe, as a
function of $x$ in La$_{2/3-x}$Tb$_x$Ca$_{1/33}$MnO$_3$ is shown in the
inset of Fig.~\ref{fig4}. The magnetization remains almost the same up
to $x$ = 0.1 and then decreases above this value of $x$. Also, the
magnetization is not saturated for $x >$ 0.1 in the measured field
range. Recent studies suggest the existence of quantum critical point
(QCP) effect in
La$_{0.67}$Ca$_{0.33}$Mn$_{1-x}$Ga$_x$O$_3$.\cite{det05} QCP is defined
as a second order transition accompanied by the change of a non-thermal
parameter. The observation of QCP in
La$_{0.67}$Ca$_{0.33}$Mn$_{1-x}$Ga$_x$O$_3$ system is as predicted by
the theoretical calculations and the QCP in this system was expected
for a value of 10-20\% of Ga substitution.\cite{alo02} The substitution
of a nonmagnetic ion like Ga at the Mn-sublattice of the
perovskite-type oxide causes the localization of the electronic states
suppressing the double exchange mechanism. For
La$_{0.67}$Ca$_{0.33}$Mn$_{1-x}$Ga$_x$O$_3$, the spontaneous magnetic
moment calculated from the experimental neutron diffraction patterns
recorded at 1.5 K decreased  for $x >$ 0.1 and vanished for $x >$ 0.16,
indicating electron localization which suppresses the double exchange
mechanism. The present dc magnetization data on
La$_{2/3-x}$Tb$_x$Ca$_{1/33}$MnO$_3$ shows almost a similar trend,
except for the finite value of the magnetization at higher values of
$x$, suggesting the possible existence of QCP in this system also.
However, this needs to be verified with the help of neutron diffraction
measurements, as made in the case of the Ga substituted system.

\begin{figure}
\includegraphics[width=3.5in,height=3in]{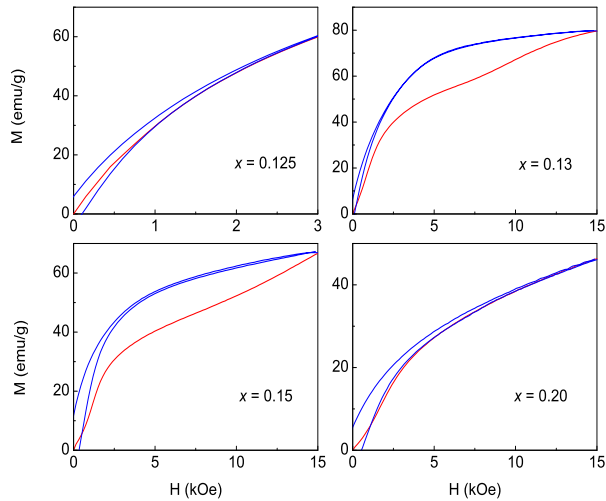}
\caption{\label{fig5} The virgin magnetization (red curves) and part
of the hysteresis loop (blue curves) of $x$ = 0.125, 0.13, 0.15, and
0.2 in La$_{0.67-x}$Tb$_x$Ca$_{0.33}$MnO$_3$.  Note that the $H$
scale is different for $x$ = 0.125.}
\end{figure}

Another interesting behavior observed in the low temperature $M-H$
measurements for the compositions immediately above $x$ = 1/8 is an
irreversible jump to a ferromagnetic state at higher magnetic fields,
as shown in Fig.~\ref{fig5}.  This is observed only for the virgin
magnetization measurements. Up to $x$ = 0.125, a normal feature is
observed, where the virgin magnetization curve lies inside the
hysteresis loop. For $x$ = 0.2, the behaviour is similar to that
observed for some typical spin glass systems,\cite{san05} where the
virgin magnetization curve initially lies outside the loop and then
merges with the loop at higher fields. On the other hand, for $x$ =
0.13 and 0.15, the entire virgin magnetization curve lies outside the
hysteresis loop above a certain small field (this small crossing field
is observed for $x$ = 0.2 and 0.25 also, and increases with $x$), and
an anomalous step-like feature is observed in the virgin magnetization
curve, similar to that of a metamagnetic transition. However, this
transition is completely irreversible. The field above which a broad
step is observed is larger for $x$ = 0.15 compared to that for $x$ =
0.13. After the magnetic field is increased in the negative direction
to -15 kOe and when brought back to +15 kOe through H = 0, the
transition is not observed. This is an irreversible ferromagnetic
transition in the sense that the step-like feature is never obtained
when the measurements were repeated immediately or even after a time
gap of 30 minutes. In the subsequent measurements, the first part of
the curve always lies inside the hysteresis loop, like that for the
compositions for $x \leq$ 0.125. It appears that an irreversible
magnetic field induced phase transition is occurred. Similar
characteristics were reported for some Pr-based substituted manganite
compositions. Dho and Hur explained this behavior in terms of the
reorientation of the Mn spins  pinned by localized Pr moments,
\cite{dho03} whereas Woodward {\it et al.}\cite{wood04} explained the
observations in terms of an avalanche behavior. It may be noted that
the first part of the virgin curve is similar to that of the spin glass
composition $x$ = 0.2, in the present case, and therefore, it is
possible that the second jump is due to a field induced growth of the
larger ferromagnetic clusters. Once the clusters are grown, it is not
possible to revert back to the original state due to the unavailability
of sufficient thermal energy. The original state is found only when the
temperature is raised above the peak temperature and then cooled back
in zero field.

\section{Conclusions}

The present studies made on a series of close compositions in
La$_{0.67-x}$Tb$_x$Ca$_{0.33}$MnO$_3$ indicate that single phase
ferromagnetic compositions are possible for $x <$ 0.1, mixed long
range ordered ferromagnetic and short range ordered magnetic
clusters coexist for 0.1 $\leq x \leq$ 0.125 and spin glass like
phases are formed for larger values of $x$. The ferromagnetic
clusters present in the compositions immediately above the cross
over region show magnetic field induced growth and give larger
magnetization at higher fields. Further detailed studies are
required to understand the complex magnetic behavior shown by
these Tb substituted manganite compositions at the intermediate
and the cross over regions.

\end{document}